\documentclass[sigconf,review=false,anonymous=false]{acmart}
\usepackage{booktabs}
\usepackage{amsmath}
\usepackage{amsfonts}
\usepackage{amssymb}
\usepackage[inline]{enumitem}
\usepackage{array}
\usepackage{censor}
\usepackage[utf8]{inputenc}
\usepackage[T1]{fontenc}
\usepackage{subcaption}
\usepackage{mathtools}
\usepackage{balance}

\copyrightyear{2018}
\acmYear{2018}
\setcopyright{acmcopyright}
\acmConference[RecSys '18]{The 12th ACM Recommender Systems Conference}{October 2--7, 2018}{Vancouver, Canada}
\acmBooktitle{RecSys '18: The 12th ACM Recommender Systems Conference, October 2--7, 2018, Vancouver, Canada}
\acmPrice{15.00}

\newcommand{\eg}{\emph{e.g.}}

\makeatletter
\g@addto@macro\normalsize{%
  \setlength\abovedisplayskip{2pt}
  \setlength\belowdisplayskip{3pt}
  \setlength\abovedisplayshortskip{2pt}
  \setlength\belowdisplayshortskip{3pt}
}

\newif\ifanon
\anonfalse

\begin{document}
\title{A Line in the Sand: Recommendation or Ad-hoc Retrieval?}
\titlenote{``A Line in the Sand'' is a reference to the song by Linkin Park from the album ``The Hunting Party'' (2014). URL: \url{https://open.spotify.com/track/4BRvD5QdauTo8EuUvYchu3?si=pPGeCtYoSLSKFVcPzexA4Q}}
\subtitle{Submission Report for RecSys 2018 Spotify Challenge by Team BachPropagate}

\author{Surya Kallumadi}
\affiliation{%
  \institution{Kansas State University}
  \city{Manhattan} 
  \state{Kansas}
  \country{USA}
}
\email{surya@ksu.edu}
\author{Bhaskar Mitra}
\affiliation{%
  \institution{Microsoft AI \& Research}
  \city{Montreal} 
  \state{Québec}
  \country{Canada}
}
\email{bmitra@microsoft.com}
\author{Tereza Iofciu}
\affiliation{%
  \institution{mytaxi}
  \city{Hamburg} 
  \country{Germany}
}
\email{t.iofciu@mytaxi.com}

\begin{abstract}
The popular approaches to recommendation and ad-hoc retrieval tasks are largely distinct in the literature.
In this work, we argue that many recommendation problems can also be cast as ad-hoc retrieval tasks.
To demonstrate this, we build a solution for the RecSys 2018 Spotify challenge by combining standard ad-hoc retrieval models and using popular retrieval tools sets.
We draw a parallel between the playlist continuation task and the task of finding good expansion terms for queries in ad-hoc retrieval, and show that standard pseudo-relevance feedback can be effective as a collaborative filtering approach.
We also use ad-hoc retrieval for content-based recommendation by treating the input playlist title as a query and associating all candidate tracks with meta-descriptions extracted from the background data.
The recommendations from these two approaches are further supplemented  by a nearest neighbor search based on track embeddings learned by a popular neural model.
Our final ranked list of recommendations is produced by a learning to rank model.
Our proposed solution using ad-hoc retrieval models achieved a competitive performance on the music recommendation task at RecSys 2018 challenge---finishing at rank 7 out of 112 participating teams and at rank 5 out of 31 teams for the main and the creative tracks, respectively.

\end{abstract}

\ccsdesc[500]{Information systems~Learning to rank}
\ccsdesc[500]{Information systems~Combination, fusion and federated search}
\ccsdesc[500]{Information systems~Recommender systems}

\keywords{Recommender systems, federated search, learning to rank}

\maketitle
\section{Introduction}
\label{sec:intro}
Recommendation and ad-hoc retrieval are two important information retrieval tasks.
Given a list of previously viewed items, a recommender system may suggest new items to the user by considering past interactions between all users and all items (\emph{collaborative filtering} \citep{breese1998empirical}), or it may suggest new items that share similar attributes to the already viewed items (\emph{content-based filtering} \citep{brusilovsky2007adaptive})---or it may adopt a \emph{hybrid} approach.
In contrast, in ad-hoc retrieval \citep{voorhees2005trec} the user expresses an explicit information need---typically in the form of a short text query---and the retrieval system responds with a ranked list of relevant information resources (e.g., documents or passages) based on the estimated match between the query and the document text.
The popular approaches to recommendation and ad-hoc retrieval tasks are largely distinct in the literature, although the two tasks share many similar properties.

The 2018 edition of the RecSys Challenge \citep{said2016short} featured the Spotify automatic playlist continuation task.
The goal is to recommend additional tracks for a playlist for which (either or both of) the title and a number of existing tracks are known.
A dataset containing one million Spotify playlists\footnote{Million Playlist Dataset, official website hosted at \url{https://recsys-challenge.spotify.com/}} is provided.
This million playlist dataset (MPD) can be used as background data, as well as for generating training examples and for offline evaluation.
Looking through the lens of a typical recommender system, we may approach this task as a collaborative filtering problem considering the playlist-track membership matrix derived from the background data.
A track may also be described by its own title, the primary artist name, the parent album name, and even the names of the playlists in which it occurs in the background data.
These descriptions can be useful for content-based filtering.
However, as the title of the paper suggests, in this work we explore how standard ad-hoc retrieval methods and tools can be useful to solve this recommendation task, using similar signals as collaborative filtering and content-based recommendation models.

We generate a collection of pseudo-documents where each document corresponds to a playlist in the background data.
The tracks in the playlist are treated as the terms in the document.
We use a standard retrieval system to index these pseudo-documents.
An input playlist---for which we should recommend new tracks---is treated as a query with its member tracks as the query terms.
Using pseduo-relevance feedback (PRF) \citep{lavrenko2008generative, lavrenko2001relevance} we generate new expansion tracks for the query and present these as our recommendations for the input playlist.
As this approach only considers past track-playlist membership information, we expect this method to recommend tracks similar to the collaborative filtering approach.

The title of the input playlist, if provided, can also be an important relevance signal.
For example, if the input playlist title is ``running music'', then tracks from other playlists titled ``running jams'' or ``running mix'' may be good candidates for recommendation.
Therefore, we create a second collection where each pseudo-document corresponds to a track in the background data.
We concatenate the titles of all the background playlists that contain the track to generate the content for these pseudo-documents.
Meta-descriptions about the track---such as, its title, its primary artist name, and its parent album name---can be similarly useful for matching against the input playlist title, and be included as part of the pseudo-documents.
We index this second collection and produce additional candidates by considering the input playlist title, if available, as a query to an ad-hoc retrieval system.

Finally, we learn track embeddings using the popular word2vec model \citep{mikolov2013efficient} and generate additional recommendations by a nearest-neighbour search in the learned latent space.
The candidates from all three approaches are combined and re-ranked using a LambdaMART model \citep{wu2010adapting}.
By using only standard IR tools and methods, we built a solution that is competitive with other top performing submissions at the RecSys 2018 Spotify Challenge.
\section{The RecSys 2018 Challenge}
\label{sec:prob}
\begin{figure}[t]
\center
\begin{subfigure}{\columnwidth}
    \includegraphics[width=0.9\linewidth]{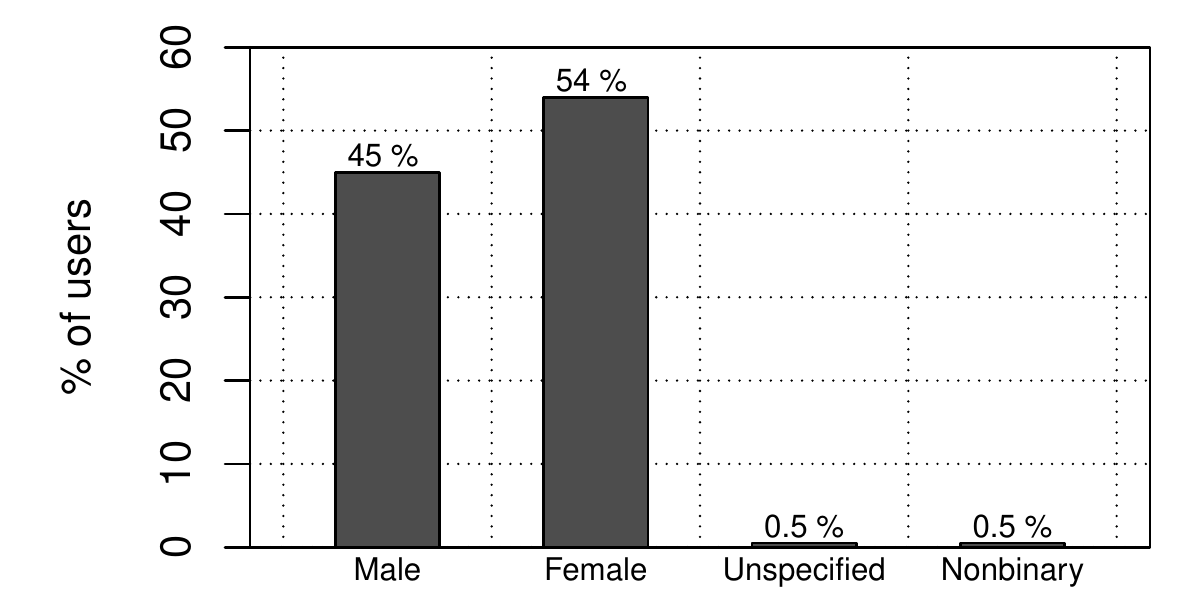}
    \caption{}
    \label{fig:demo-gender}
\end{subfigure}
\hfill
\begin{subfigure}{\columnwidth}
    \includegraphics[width=0.9\linewidth]{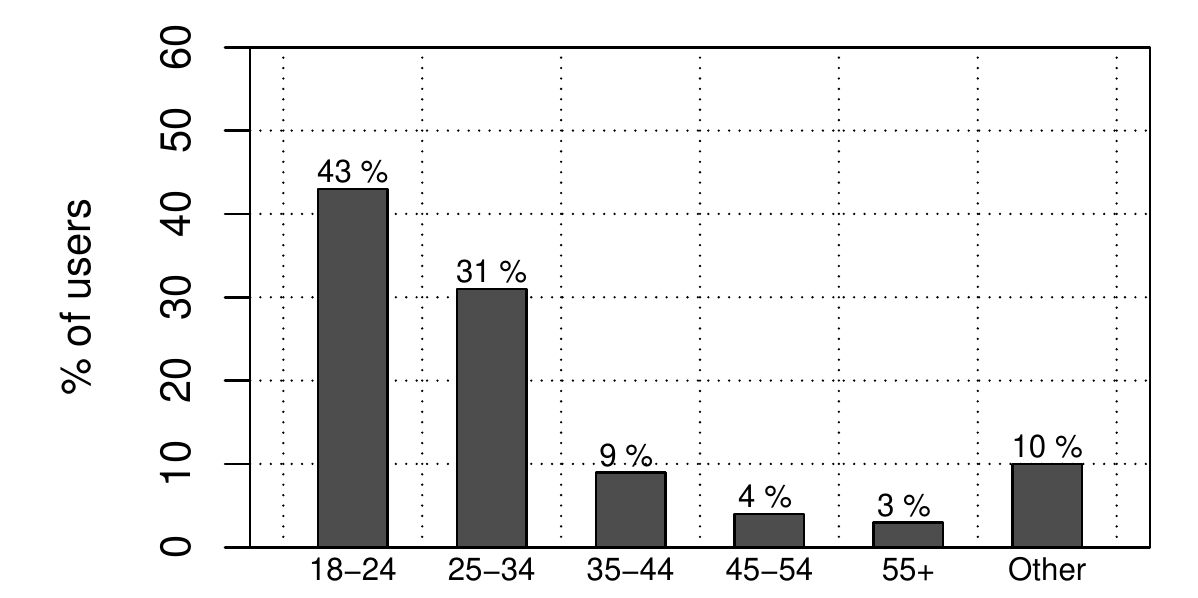}
    \caption{}
    \label{fig:demo-age}
\end{subfigure}
\caption{Demographics of users who contributed to the MPD by (a) gender and (b) age.}
\label{fig:data-demo}
\end{figure}

Spotify---an online music streaming company\footnote{\url{https://www.spotify.com/}}---co-organized the RecSys 2018 challenge.
The goal of this year's challenge was music recommendation---to suggest new tracks for playlist continuation.
As part of this challenge, Spotify released a dataset containing one million randomly sampled user generated playlists that are publicly available to any users of the music streaming platform.
The dataset includes playlists that were created between January 1, 2010 and November 1, 2017 by users who are at least 13 years old and resident in the United States.
Any private user information is excluded from the dataset, and adult and offensive content scrubbed.
Additional constraints placed on the inclusion of any playlist in this dataset include:
\begin{enumerate*}[label=(\roman*)]
  \item a minimum number other playlists that should contain the same title,
  \item a minimum of three distinct artists and two distinct albums in the playlist,
  \item at least one follower other than the creator, and
  \item no less than five and no more than 250 tracks in the playlist.
\end{enumerate*}
The demographic distribution of the users who contributed to the dataset---according to the challenge website\footnote{\url{https://recsys-challenge.spotify.com/dataset}}---is reproduced in Figure~\ref{fig:data-demo}.

The challenge dataset contains ten thousand playlists.
For each playlist $\Phi = \phi_\text{seed} \cup \phi_\text{held}$, a set of tracks $\phi_\text{seed} = \{{tr}_1, {tr}_2, \ldots, {tr}_m\}$ are provided as seed tracks and the remaining tracks $\phi_\text{held} = \{{tr}_1, {tr}_2, \ldots, {tr}_n\}$ have been heldout.
Optionally, the title $T_\Phi$ of the playlist $\Phi$ is also provided.
The recommendation task involves predicting the heldout tracks in $\phi_\text{held}$ given $\phi_\text{seed}$ and optionally $T_\Phi$.
The number of heldout tracks $n$ for each playlist $\Phi$ is known and each playlist in the challenge set belongs to one of the following ten categories based on the information provided.
\begin{enumerate*}[label=(\roman*)]
  \item the title only,
  \item the title and the first track,
  \item the title and the first five tracks,
  \item the first five tracks only,
  \item the title and the first ten tracks,
  \item the first ten tracks only,
  \item the title and the first 25 tracks,
  \item the title and 25 random tracks,
  \item the title and the first 100 tracks, and
  \item the title and 100 random tracks.
\end{enumerate*}
When track information is provided, each track ${tr}$ is described by:
\begin{enumerate*}[label=(\roman*)]
  \item its position in the playlist,
  \item the track name,
  \item the track URI,
  \item the primary artist name,
  \item the primary artist URI,
  \item the album name,
  \item the album URI, and
  \item its duration.
\end{enumerate*}
The challenge set is sampled following the same guidelines as the MPD.
For each playlist, the recommender system needs to generate a ranked list of exactly 500 distinct tracks $\phi_\text{pred}$ with no overlap with the seed tracks $\phi_\text{seed}$ provided as part of the playlist information.

Submissions are accepted under two different tracks---the main track and the creative track.
For the creative track, participants are allowed to use external data for making the recommendations.
The use of external data, however, is restricted to those that are publicly available to all participants.

Each submission is evaluated based on three different metrics:
\begin{enumerate}
  \item R-precision \citep{voorhees2003common}, with partial credit for artist match even if the track is incorrect
  \item Normalized Discounted Cumulative Gain (NDCG) \citep{jarvelin2002cumulated}
  \item Recommended songs clicks, computed as:
  \begin{align}
  clicks = min\{\lfloor (r - 1) / 10 \rfloor, 51\}
  \end{align}
  where, $r$ is the highest rank of a relevant track, if any.
\end{enumerate}

The challenge leaderboard ranked each participants based on the Borda Count \citep{de1781memoire} election strategy over all the three specified metrics.
During the submission stage, the leaderboard reflected the ranking based on a fixed 50\% random sample of the actual challenge set.
However, at the end of the competition the final ranking was computed based on the full set.
For more details, we point the readers to the official rules as listed on the challenge website: \url{https://recsys-challenge.spotify.com/rules}.
\section{Our approach}
\label{sec:method}
Our proposed solution consists of a candidate generation stage and a re-ranking stage. To recall a diverse set of candidates for ranking, we employ three different candidate generation strategies.
Two of these approaches depend on track co-occurrence information, and the other approach models the relationship between tracks and the titles of parent playlists.
Two of the approaches are implemented using Indri\footnote{\url{http://www.lemurproject.org/indri/}}---a standard ad-hoc retrieval system---while the other employs a nearest neighbor based lookup.
We describe all three candidate generation methods and the re-ranking model next.

\begin{table*}
\caption{The full list of features that our learning to rank model considers. The features are categorized based on whether they depend only on the input playlist or the candidate track, or both.}
\label{tbl:method-ltr-features}
\begin{center}
\resizebox{\textwidth}{!}{
\setlength{\tabcolsep}{4pt}
\begin{tabular}{l l}
\toprule
Features & Types \\
\midrule
\multicolumn{2}{l}{\textbf{Input playlist only features}} \\
Is playlist title available & Binary \\
Number of total tracks & Integer \\
Number of held out tracks & Integer \\
Ratio of number of unique albums to number of tracks & Float \\
Ratio of number of unique artists to number of tracks & Float \\
Ratio of frequency of most frequent album to number of tracks & Float \\
Ratio of frequency of most frequent artist to number of tracks & Float \\
Playlist title contains any of the words: top, best, popular, hot, or hits & Binary \\
Playlist title contains any of the words: latest, new, or recent & Binary \\
Playlist title contains any of the words: remix, remixed, or remixes & Binary \\
\midrule
\multicolumn{2}{l}{\textbf{Candidate track only features}} \\
Ratio of number of background playlists containing this track to total number of background playlists & Float \\
Ratio of number of background playlists containing this artist to total number of background playlists & Float \\
Ratio of number of background playlists containing this album to total number of background playlists & Float \\
Track title contains any of the words: remix, remixed, or remixes & Binary \\
Ratio of number of background parent playlists with title containinng any of the words: top, best, popular, hot, or hits to total number of background playlists & Float \\
Ratio of number of background parent playlists with title containing any of the words: lates, new, or recent to total number of background playlists & Float \\
Ratio of number of background parent playlists with title containing any of the words: remix, remixed, or remixes to total number of background playlists & Float \\
\midrule
\multicolumn{2}{l}{\textbf{Input playlist and candidate track dependent features}} \\
Rank in top 1000 candidates from QE, set to 1001 if not present & Integer \\
Rank in top 500 candidates from META1, set to 501 if not present & Integer \\
Rank in top 500 candidates from META2, set to 501 if not present & Integer \\
Rank in top 250 candidates from EMB1, set to 251 if not present & Integer \\
Rank in top 250 candidates from EMB2, set to 251 if not present & Integer \\
Rank in top 250 candidates from EMB3, set to 251 if not present & Integer \\
Rank in top 250 candidates from EMB4, set to 251 if not present & Integer \\
Ratio of number of tracks in playlist from same artist to number of tracks in playlist & Float \\
Ratio of number of tracks in playlist from same album to number of tracks in playlist & Float \\
\bottomrule
\end{tabular}
}
\end{center}
\end{table*}

\subsection{Candidate generation}
\label{sec:method-candgen}

\paragraph{Playlist completion as query expansion (QE)}
In PRF \citep{lavrenko2008generative, lavrenko2001relevance}, given a query $q$ of $m$ terms $\{t_1, t_2, \ldots, t_m\}$, first a set of $k$ documents $D = \{d_1, d_2, \ldots,d_k\}$ are retrieved and based on these retrieved documents $D$ the query is updated to $q'$. 
The translation from $q$ to $q'$ typically involves addition of new terms from $D$ to the original query $q$.
A new round of retrieval is performed using $q'$ and the newly retrieved documents presented to the user.

Let us consider individual tracks as terms and playlists as text---like a document or a query---containing one or more terms.
Let us also assume that we have an incomplete playlist $\phi_\text{seed}$ which is derived from an original playlist $\Phi$.
Let $C$ be the collection of all playlists in the MPD and let $C' = C \cup \{\Phi\}$ be an imaginary collection created by adding $\Phi$ to $C$.
Now, say, we want to retrieve $\Phi$ from $C'$ but we are only provided $\phi_\text{seed}$ as a query.
We know that we can obtain a smoother estimate of the unigram distribution of terms (or tracks) in $\Phi$---and hence a better retrieval performance on this retrieval task---by first expanding $\phi_\text{seed}$ to $\phi_\text{exp} = \phi_\text{seed} \cup \phi_\text{new}$, where $\phi_\text{new}$ is the set of additional ``query terms'' identified by performing PRF over the collection $C$.
While we do not, in fact, have $C'$ and nor are we interested in retrieving $\Phi$ from this imaginary collection, it is interesting to note that PRF over $C$ starting from $\phi_\text{seed}$ can help us identify a set of terms (or tracks) that are potentially from $\Phi$ but missing in $\phi_\text{seed}$.
Estimating $\phi_\text{new}$ accurately is similar to our playlist completion task.
We note that a similar approach has been previously proposed for collaborative filtering \citep{parapar2013relevance, valcarce2016efficient}.

Motivated by this, we use Indri to index a collection of all the the playlists in the MPD, where each playlist is a sequence of track identifiers.
Given an incomplete playlist $\phi_\text{seed}$, we retrieve a set of k playlists $c$ from the collection and identify good expansion terms (or tracks) using RM1 \citep{abdul2004umass}.

\begin{align}
p(tr|\theta_\Phi) &= \sum_{\phi \in c}{p(tr|\theta_\phi)\prod_{\bar{tr} \in \phi_\text{seed}}{p(\bar{tr}|\theta_\phi)}} \\
p(tr|\theta_\phi) &= \frac{|\phi \cap \{tr\}|}{|\phi|},\qquad\text{without smoothing}
\end{align}

The top candidate tracks ranked by $p(tr|\theta_\Phi)$ are considered for recommendation.
We refer to this candidate generation strategy as QE in the rest of this paper.

\paragraph*{Ad-hoc track retrieval using meta descriptions (META)}
In ad-hoc retrieval, a document representation may depend on its own content---\eg, title or body text--- or external sources of descriptions---\eg, anchor text or clicked queries \citep{zamani2018neural, robertson2004simple}.
Similarly, we can describe a track by its own title, the primary artist name, and the parent album name---or by the titles of all the playlists in which it appears.
All of these meta information about the track may be useful for our recommendation task.
Given an input playlist title $T_\Phi$, we can query a collection of pseudo-documents---where each document contains meta descriptions for a track---using a standard retrieval system, such as Indri.
The retrieved ranked list of tracks can be considered as candidates for the playlist completion task.
Based on this intuition, we generate two collections---one that describes tracks by their parent playlist titles and another that describes a track by its own title, primary artist name, and album name. Separate set of candidates retrieved based on each of these two collections are referred to as META1 and META2, respectively, in the rest of this paper.
In our specific implementation, we use BM25 \citep{robertson2009probabilistic} as the retrieval model and each document is generated by concatenation of the constituent text descriptions, similar to \citet{robertson2004simple}.

\paragraph*{Nearest neighbor search using track embeddings (EMB)}
Instead of comparing the query and the document text in the term space, some ad-hoc retrieval models---\eg, \citep{le2014distributed, mitra2016desm, zamani2016estimating}---compute the query and the document representations as a centroid of their term embeddings and estimate their similarity in the latent space.
A similar strategy may be useful for the playlist completion task.
We experiment with a number of unsupervised approaches to learn the track embeddings that do not require any additional manual annotations.

First, we consider tracks as terms and playlists as documents containing a sequence of tracks.
We employ the popular CBOW model from word2vec \citep{mikolov2013efficient} to learn track embeddings on this pseduo-document collection $C$.
A fixed size window is moved over each playlist and the model is trained by trying to predict the track in the middle of the window correctly given all the other tracks within the same window.
This translates to minimizing the following loss,

\begin{align}
\mathcal{L}_\text{CBOW} = \sum_{\phi \in C}\sum_{i}^{|\phi|}{-\text{log}\big(p(\vec{v}_{i}|\sum_{i-k \leq j \leq i+k, j \neq i}{\vec{v}_{j}})\big)  }
\end{align}

Where, $\vec{v}_{i}$ is the embedding of the $i^\text{th}$ track in the playlist $\phi$.
Similar to \citet{mikolov2013distributed}, our track embeddings are trained with negative sampling instead of the full softmax over the complete track collection.

A playlist representation can be derived from both its member tracks $\{{tr}_1,{tr}_2,\ldots,{tr}_m,\}$ as well as its title $T_\phi$.
An analogy can be drawn to two collections in two different languages with document aligned across the collections.
\citet{vulic2015monolingual} consider a similar scenario in the context of cross-lingual retrieval and propose to learn a shared embedding space for terms from both languages by merging the two versions of each document from respective languages into a single pseudo-document.
Motivated by their approach, we generate a collection of playlists where each pseudo-document is constructed by interspersing the member tracks and the playlist title terms.
We train a CBOW model on this collection as our second approach to learn track embeddings.

The MPD contains four different types of entities---playlists, tracks, artists, and albums.
Alternatively, we can view this dataset as a Heterogeneous Information Network (HIN).
A HIN is defined as a directed graph $G = \{V, E\}$ with an entity mapping function $\xi : \mathbb V\to\mathbb \mathcal{A}$ and a edge type mapping function $\psi : \mathbb E\to\mathbb \mathcal{R}$ where each node $ $v $\in  \mathbb V $ belongs to one particular entity type  $\xi($v$) \in \mathcal{A}$ and each edge $ $e $\in  \mathbb E $ belongs to a relationship type $\psi($e$) \in \mathcal{R}$. The edge weights associated between vertices with the relationship context $\psi($c$) \in \mathcal{R}$ is captured as a preference matrix $\mathcal{W}_c$.
Finally, a meta-path defines a composite relationship by an ordered sequence of edge types specified in the HIN schema $\mathcal{S}_G = (\mathcal{A},\mathcal{R})$.
A number of previous studies have explored methods to learn node embeddings in homogeneous \citep{perozzi2014deepwalk, grover2016node2vec, tang2015line} and heterogeneous \citep{dong2017metapath2vec, tang2015pte} graphs.
In particular, \citet{dong2017metapath2vec} propose meta-path based random walks in heterogeneous networks to generate neighborhood representations that capture semantic relationships between different types of nodes in the graph followed by training a word2vec model on this neighborhood data to learn node embeddings.
We adopt a similar approach based on two different meta-path definitions: artist$\rightarrow$track$\rightarrow$playlist$\rightarrow$artist (ATPA) and track$\rightarrow$playlist$\rightarrow$track (TPT).
In summary, we learn track embeddings based on four different approaches:
\begin{itemize}
  \item EMB1: CBOW over playlists as documents and tracks as terms
  \item EMB2: CBOW over interspersed member tracks and title terms for a playlist
  \item EMB3: CBOW over the ATPA meta path
  \item EMB4: CBOW over the TPT meta path
\end{itemize}

After training, we represent an input playlist $\phi_\text{seed}$ as the average of its member track embeddings $\vec{v}_\text{seed}$.
New recommendation candidates are identified by finding tracks that have high cosine similarity with $\vec{v}_\text{seed}$.
The embedding size is fixed to 200 dimensions for all four approaches and the window size for word2vec at 20 for EMB1 and EMB2 and at 5 for EMB3 and EMB4.

\subsection{Learning to rank}
\label{sec:method-ranking}

\begin{table*}
\caption{Offline evaluation results for individual candidate sources and the combined LTR model output. For the combined model, we only measured the metrics at rank 500. The combined model achieves the best performance while QE emerges as the best candidate source. Note that for the clicks metric a lower value indicates a better performance.}
\label{tbl:results-offline}
\begin{center}
\resizebox{\textwidth}{!}{
\setlength{\tabcolsep}{4pt}
\begin{tabular}{l | l l l l l l l l l l l l l l l l l}
\toprule
 & & \multicolumn{4}{c}{Recall} & & \multicolumn{4}{c}{RPrec} & & \multicolumn{4}{c}{NDCG} & & \multicolumn{1}{c}{Clicks} \\ \cline{3-6} \cline{8-11} \cline{13-16}
Model & & @10 & @250 & @500 & @1000 & & @10 & @250 & @500 & @1000 & & @10 & @250 & @500 & @1000 & & @500 \\
\midrule
QE & & 0.072 & 0.392 & 0.497 & 0.596 & & 0.063 & 0.129 & 0.129 & 0.129 & & 0.204 & 0.264 & 0.303 & 0.337 & & 05.129 \\
META1 & & 0.033 & 0.232 & 0.309 & 0.393 & & 0.032 & 0.100 & 0.100 & 0.100 & & 0.160 & 0.181 & 0.217 & 0.252 & & 08.839 \\
META2 & & 0.001 & 0.012 & 0.016 & 0.018 & & 0.001 & 0.003 & 0.003 & 0.003 & & 0.003 & 0.007 & 0.009 & 0.010 & & 47.857 \\
EMB1 & & 0.025 & 0.129 & 0.174 & 0.234 & & 0.022 & 0.038 & 0.038 & 0.038 & & 0.065 & 0.084 & 0.099 & 0.118 & & 21.740 \\
EMB2 & & 0.031 & 0.156 & 0.200 & 0.250 & & 0.028 & 0.049 & 0.049 & 0.049 & & 0.087 & 0.104 & 0.119 & 0.135 & & 17.531 \\
EMB3 & & 0.042 & 0.174 & 0.214 & 0.261 & & 0.038 & 0.065 & 0.065 & 0.065 & & 0.116 & 0.126 & 0.140 & 0.155 & & 21.112 \\
EMB4 & & 0.048 & 0.219 & 0.268 & 0.320 & & 0.043 & 0.078 & 0.078 & 0.078 & & 0.138 & 0.155 & 0.173 & 0.190 & & 17.174 \\
All candidate sources + LTR & & - & - & 0.513 & - & & - & - & 0.134 & - & & - & - & 0.313 & - & & 04.380 \\
\bottomrule
\end{tabular}
}
\end{center}
\end{table*}

We take the union of all the candidates generated by each of the approaches described in Section \ref{sec:method-candgen}.
More precisely, we take the top 1000 candidates from QE, top 500 candidates each from META1 and META2, and top 250 candidates each from EMB1, EMB2, EMB3, and EMB4.
We re-rank these candidates using a learning to rank (LTR) \citep{Liu:2009} model.
We choose LambdaMART \citep{wu2010adapting} with 100 trees and 50 leaves per tree as our model.
We use the publicly available implementation in RankLib\footnote{\url{https://sourceforge.net/p/lemur/wiki/RankLib/}} for our experiments.
We train the model with a learning rate of 0.1 and optimize for NDCG@10 for our main track submission and for NDCG@500 for our submission to the creative track.
The full list of features used by the LTR model is specified in Table \ref{tbl:method-ltr-features}.

During the LTR model training, we use 75\% of the MPD for candidate generation and feature computation.
From the remaining portion we use 50K playlists to train the LTR model and 5K playlists for offline evaluation.
For each playlist in both the train and the evaluation, we hold out some of the member tracks---and optionally the playlist title---to generate a dataset with similar distributions as the challenge set.
After finalizing the LTR model, we regenerate the candidates and recompute the features using the full MPD for the final challenge submission.

\smallskip\noindent
An open source implementation of our framework is available at \url{https://github.com/skallumadi/BachPropagate}.
\section{Results}
\label{sec:result}

Table \ref{tbl:results-offline} shows the offline evaluation results for the individual candidate generation strategies and the final combined output of the LTR model.
Among the different candidate sources, QE demonstrates the strongest performance across all four metrics and all rank positions.
While META1 shows reasonable performance, META2 achieves modest results most likely because the challenge set is designed such that each playlist containts a diverse set of artists and albums.
So matching the input playlist title with the candidate track's title or its album/artist name does not add enough value.
EMB4 fares the best among all the track embedding based approaches.
The LTR model that re-ranks a combined set of candidates from all the different sources performs best and shows significant improvement over the strongest individual source QE.

The final standing on the RecSys 2018 challenge for the main and the creative tracks are shown in Table \ref{tbl:results-leaderboard}.
Our submission based on the framework described in this paper features among the top ten teams out of 112 participants on the main track and among the top five teams out of 31 teams on the creative track.
Our submission also ranked among the top five teams based on the clicks metric alone on both tracks.
We achieved this competitive performance based on simple applications of standard IR models.
Our approach may be improved even further by incorporating more advanced retrieval models, including those based on recent neural and other machine learning based approaches \citep{mitra2017introduction}.

\begin{table}
\caption{The final RecSys 2018 spotify challenge leaderboards. Our submissions are highlighted in bold. Only the top 10 teams from the leaderboards are shown. The total number of participating teams was 112 and 31 for the main and the creative tracks, respectively. For the clicks metric a lower value indicates a better performance.}
\label{tbl:results-leaderboard}
\begin{center}
\begin{subtable}[a]{\columnwidth}
\resizebox{\columnwidth}{!}{
\setlength{\tabcolsep}{4pt}
\begin{tabular}{l l c c c c c c c}
\toprule
& & \multicolumn{2}{c}{RPrec} & \multicolumn{2}{c}{NDCG} & \multicolumn{2}{c}{Clicks} & \\ 
\# & Team name & Value & Rank & Value & Rank & Value & Rank & Borda \\
\midrule
1 & vl6 & 0.224 & 1 & 0.395 & 1 & 1.784 & 2 & 329 \\
2 & hello world\! & 0.223 & 2 & 0.393 & 2 & 1.895 & 6 & 323 \\
3 & Avito & 0.215 & 6 & 0.385 & 4 & 1.782 & 1 & 322 \\
4 & Creamy Fireflies & 0.220 & 3 & 0.386 & 3 & 1.934 & 7 & 320 \\
4 & MIPT\_MSU & 0.217 & 4 & 0.382 & 5 & 1.875 & 4 & 320 \\
6 & HAIR & 0.216 & 5 & 0.380 & 6 & 2.182 & 13 & 309 \\
7 & KAENEN & 0.209 & 11 & 0.375 & 8 & 2.054 & 10 & 304 \\
\textbf{7} & \textbf{BachPropagate} & \textbf{0.209} & \textbf{12} & \textbf{0.374} & \textbf{12} & \textbf{1.883} & \textbf{5} & \textbf{304} \\
9 & Definitive Turtles & 0.209 & 13 & 0.375 & 7 & 2.078 & 11 & 302 \\
10 & IN3PD & 0.208 & 14 & 0.371 & 14 & 1.952 & 8 & 297 \\
\bottomrule
\end{tabular}
}
\vspace{.3\baselineskip}
\subcaption{Main track}
\vspace{.5\baselineskip}
\end{subtable}
\begin{subtable}[b]{\columnwidth}
\resizebox{\columnwidth}{!}{
\setlength{\tabcolsep}{4pt}
\begin{tabular}{l l c c c c c c c}
\toprule
& & \multicolumn{2}{c}{RPrec} & \multicolumn{2}{c}{NDCG} & \multicolumn{2}{c}{Clicks} & \\ 
\# & Team name & Value & Rank & Value & Rank & Value & Rank & Borda \\
\midrule
1 & vl6 & 0.223 & 1 & 0.394 & 1 & 1.785 & 1 & 90 \\
2 & Creamy Fireflies & 0.220 & 2 & 0.385 & 2 & 1.925 & 4 & 85 \\
3 & KAENEN & 0.209 & 3 & 0.375 & 3 & 2.048 & 6 & 81 \\
4 & cocoplaya & 0.202 & 7 & 0.366 & 6 & 1.838 & 2 & 78 \\
\textbf{5} & \textbf{BachPropagate} & \textbf{0.202} & \textbf{6} & \textbf{0.366} & \textbf{5} & \textbf{2.003} & \textbf{5} & \textbf{77} \\
6 & Trailmix & 0.206 & 4 & 0.370 & 4 & 2.259 & 9 & 76 \\
7 & teamrozik & 0.205 & 5 & 0.361 & 7 & 2.164 & 8 & 73 \\
8 & Freshwater Sea & 0.195 & 9 & 0.350 & 9 & 2.130 & 7 & 68 \\
9 & Team Radboud & 0.198 & 8 & 0.356 & 8 & 2.293 & 11 & 66 \\
10 & spotif.ai & 0.192 & 10 & 0.339 & 11 & 2.267 & 10 & 62 \\
\bottomrule
\end{tabular}
}
\vspace{.3\baselineskip}
\subcaption{Creative track}
\end{subtable}
\end{center}
\end{table}

\section{Conclusion}
\label{sec:conclusion}
In this paper, we have argued that ad-hoc retrieval models can be useful for recommendation tasks.
However, so far we have based our argument solely on retrieval performance.
Another important consideration in this debate is the runtime efficiency.
Using inverted index and other specialized data structures, typical web scale IR systems can retrieve the relevant results under a second from collections containing more than billions of items \citep{teevan2013slow}.
The Recsys 2018 challenge does not consider runtime efficiency.
It is likely that our argument for applying ad-hoc retrieval models to recommendation tasks may be strengthened if we consider model response times.

Finally, because our main goal in this work was to achieve a competitive performance at this year's RecSys challenge, the current study is focused primarily on empirical results.
However, a theoretical comparison of ad-hoc retrieval models and recommender systems may reveal more insights and opportunities in the intersection of these two research communities.
We conclude by highlighting this as an important direction for future work in this area.


\bibliographystyle{ACM-Reference-Format}
\balance
\bibliography{bibtex.bib}

\end{document}